\documentclass[11pt]{article} 
\usepackage{latex8}

\newtheorem{theorem}{Theorem}[section]
\newtheorem{lemma}[theorem]{Lemma}
\newtheorem{fact}[theorem]{Fact}

\newtheorem{claim}[theorem]{Claim}

\def\square{\rule{2mm}{2mm}}

\newenvironment{proof}{{\noindent\bf Proof:  }}{\qquad\square}
\newenvironment{proofof}[1]{{\noindent\bf Proof of #1:  }}{\qquad\square}

\def\squarebox#1{\hbox to #1{\hfill\vbox to #1{\vfill}}}

\newcommand{\tensor}{\otimes}

\newcommand\meet\wedge
\newcommand\implies\Rightarrow

\def\ccc{{\mathchoice {\setbox0=\hbox{$\displaystyle\rm
C$}\hbox{\hbox
to0pt{\kern0.4\wd0\vrule height0.9\ht0\hss}\box0}}
{\setbox0=\hbox{$\textstyle\rm C$}\hbox{\hbox
to0pt{\kern0.4\wd0\vrule height0.9\ht0\hss}\box0}}
{\setbox0=\hbox{$\scriptstyle\rm C$}\hbox{\hbox
to0pt{\kern0.4\wd0\vrule height0.9\ht0\hss}\box0}}
{\setbox0=\hbox{$\scriptscriptstyle\rm C$}\hbox{\hbox
to0pt{\kern0.4\wd0\vrule height0.9\ht0\hss}\box0}}}}

\newcommand{\reals}{{\hbox{\sf I\kern-.14em\hbox{R}}}}
\newcommand{\trace}{{\rm Tr}}
\newcommand{\prob}{{\rm Prob}}

\newcommand{\size}[1]{\left|#1\right|}

\newcommand{\ket}[1]{\left|#1\right\rangle}
\newcommand{\bra}[1]{\left\langle #1\right|}
\newcommand{\braket}[2]{\left\langle #1\!\mid\! #2\right\rangle}
\newcommand{\ketbra}[2]{\ket{#1}\!\bra{#2}}
\newcommand{\norm}[1]{\left\|\,#1\,\right\|}
\newcommand{\trnorm}[1]{\norm{#1}_{\rm t}}
\newcommand{\set}[1]{{\left\{#1\right\}}}

\newcommand{\remove}[1]{}

\newcommand{\co}{{\cal O}}
\newcommand{\ch}{{\cal H}}
\newcommand{\disj}{{\rm DISJ}}

\newcommand{\aitch}{{\mathcal H}}
\newcommand{\kay}{{\mathcal K}}
\newcommand{\bsigma}{{\mathbf \sigma}}
\newcommand{\brho}{{\mathbf \rho}}

\newcommand{\trn}[1]{\trnorm{#1}}
\newcommand{\la}{\langle}
\newcommand{\ra}{\rangle}
\newcommand{\density}[1]{\ketbra{#1}{#1}}

\newcommand{\X}{{\mathcal X}}
\newcommand{\Y}{{\mathcal Y}}
\newcommand{\cp}{{\mathcal P}}

\begin{document}

\title{\Large\bf Interaction in Quantum Communication Complexity}

\author{    
     Ashwin Nayak
\thanks{Supported by a joint DIMACS-AT{\&}T
             Post-Doctoral Fellowship. Part of this work was completed while
            the author was at University of California, Berkeley, and
            was supported by a JSEP grant and NSF grant CCR~9800024.} \\
     DIMACS Center \\
     Rutgers, P.O.\ Box~1179  \\
     Piscataway, NJ 08855 \\
     nayak@dimacs.rutgers.edu
     \and
     Amnon Ta-Shma 
\thanks{Supported in part by a David and Lucile 
 Packard Fellowship for Science and Engineering and
 NSF NYI Grant No.\ CCR-9457799.}\\
     Computer Science Division \\
     University of California\\
     Berkeley, CA 94720 \\
     amnon@cs.berkeley.edu
     \and
     David Zuckerman
\thanks{On leave from the University of Texas at Austin.
 Supported in part by a David and Lucile 
 Packard Fellowship for Science and Engineering,
 NSF NYI Grant No.\ CCR-9457799,
 and an Alfred P.\ Sloan Research Fellowship.}\\
     Computer Science Division \\
     University of California\\
     Berkeley, CA 94720 \\
     diz@cs.berkeley.edu
}

\maketitle

\begin{abstract}
One of the most intriguing facts about communication using quantum
states is that these states cannot be used to transmit more classical bits 
than the number of qubits used, yet there are ways of conveying information 
with exponentially fewer qubits than possible classically~\cite{ASTVW98,R99}.
Moreover, these methods
have a very simple structure---they involve little interaction between
the communicating parties.
We look more closely at
the ways in which information encoded in quantum states may be
manipulated, and consider the question as to whether every classical
protocol may be transformed to a ``simpler'' quantum protocol of similar
efficiency. By a simpler protocol,
we mean a protocol that uses fewer message exchanges.
We show that for any constant~$k$, there is a problem such that
its~$k+1$ message classical communication complexity 
is exponentially smaller than its~$k$
message quantum communication complexity, thus answering the above 
question in the negative. 
Our result builds on two primitives, 
{\em local transitions in bi-partite states} (based on previous work)
and {\em average encoding\/} 
which may
be of significance in other applications as well.
\end{abstract}

\section{Introduction}
\label{sec-intro}


A recurring theme in quantum information processing has been the idea of
exploiting the exponential resources afforded by quantum states to
encode information in very non-obvious ways. Perhaps the most
representative result of this kind is due to Ambainis, Schulman,
Ta-Shma, Vazirani and Wigderson~\cite{ASTVW98}, which shows
that it is possible to deal a random set of~$\sqrt{N}$ cards each from a set
of~$N$ by the exchange of~$O(\log N)$ quantum bits between two players.
Raz~\cite{R99} gives a communication problem where the
information storage capacity of quantum states is exploited more
explicitly. Both are examples of problems for which exponentially fewer
quantum bits are required to accomplish a communication
task, as compared to classical
bits.

The protocols presented by~\cite{ASTVW98,R99} also share the feature
that they require minimal {\em interaction\/} 
between the communicating players. For example, in the
first protocol, one player prepares a set of qubits in a certain 
state and sends half of them across as the message,
after which both players measure
their qubits to obtain the result. On the other hand, efficient quantum
protocols for problems such as checking set disjointness (\disj)
seem to require much more interaction:
Buhrman, Cleve and Wigderson~\cite{bcw} give an~$O(\sqrt{N}\log N)$ qubit
protocol for \disj\ that takes~$O(\sqrt{N})$ message exchanges. This 
represents quadratic savings in communication cost,
but also an {\em unbounded\/} increase
in the number of messages exchanged (from one message to~$\sqrt{N}$),
as compared to
classical protocols. Can we
exploit the features of quantum communication and always
reduce interaction while maintaining the same communication cost? In
other words, do all efficient quantum protocols have the simple
structure shared by those of~\cite{ASTVW98,R99}?

In this paper, we study the effect of interaction on the quantum
communication complexity of problems. We show that
for any constant~$k$, allowing even {\em one\/} more message may lead to 
an exponential decrease in the communication complexity of a problem,
thus answering the above question in the negative. More formally,
\begin{theorem}
\label{thm-main}
For any constant~$k$, there is a problem
such that 
any quantum protocol with only~$k$ messages and constant probability of
error requires~$\Omega(N^{1/(k+1)})$ communication
qubits, whereas it
can be solved with~$k+1$ messages by a deterministic protocol
with~$O(\log N)$ bits.
\end{theorem}
Klauck~\cite{Klauck00} states a relationship between the bounded message
complexity of Pointer Jumping and \disj. Together with our result, this
implies an~$\Omega(N^{1/k(k+1)})$ lower bound for~$k$ message
protocols for \disj, for any constant~$k$.


The role of interaction in classical communication is well-studied,
especially in the context of the pointer jumping
function~\cite{PS82,DGS87,NW93,PRV99}.
In fact, the problem we study in this paper 
is the subproblem of Pointer Jumping singled out in~\cite{MNSW95}. 
Our analysis has the same gross structure as that in~\cite{MNSW95} (also
explained in~\cite{KN97}), but relies on entirely new ideas from quantum
information theory.

In the context of quantum communication, it was
observed by Buhrman and de Wolf~\cite{BW99} (based on a lower bound of
Nayak~\cite{Nayak99}) that any one message quantum
protocol for \disj\ has linear communication complexity. Thus, allowing more
interaction leads to a quadratic improvement in communication cost.
The lower bound of~\cite{Nayak99} immediately implies a much stronger
separation: it shows that the two message complexity of a problem may be
exponentially smaller than its one message complexity (see
also~\cite{Klauck00}). Our result subsumes all these.


Our interest in the role of interaction in quantum communication also
springs from the need to better understand the ways in which we can
access and manipulate information encoded in quantum states. We
develop information-theoretic techniques that expose some of the
limitations of quantum communication. More specifically, we present 
a new primitive in quantum encoding, as suggested by the following
theorem.
\begin{theorem}[Average encoding theorem]
\label{thm-mi}
Let~$x \mapsto \bsigma_x$ be a quantum encoding
mapping~$m$ bit strings~$x\in \set{0,1}^m$ into mixed
states~$\bsigma_x$.
Let~$X$ be distributed uniformly over~$\set{0,1}^m$, let~$Q$ be the
encoding of~$X$ according to this map, and let~$\bsigma = {1\over{2^m}} 
\sum_x \bsigma_x$.
Then,
\begin{eqnarray*}
\frac{1}{2^m} \sum_x \trn{\bsigma - \bsigma_x} & \le & 2\sqrt{I(Q:X)}.
\end{eqnarray*}
\end{theorem}
In other words, if an encoding~$Q$ is only weakly correlated to a random
variable~$X$, then the ``average encoding''~$\bsigma$ (corresponding to a
random string)
is on average a good approximation of any encoded state. Thus, in certain
situations, we may dispense with the encoding altogether, and use the single
state~$\bsigma$ instead.

We also use another primitive derived from the work of
Lo and Chau~\cite{LC98} and Mayers~\cite{M97}
which combines results of 
Jozsa~\cite{Jozsa94}, and Fuchs and van de Graaf~\cite{FG99}.
Consider two bi-partite pure
states such that one party sharing the states cannot
locally distinguish well between the two states with good probability. 
Then the other party can locally transform any of the states close to
the other.
\begin{theorem}[Local transition theorem]
(based on \cite{LC98,M97,Jozsa94,FG99})
\label{thm:close}
Let~$\brho_1,\brho_2$ be two mixed states with support in a  Hilbert
space $\aitch$,
$\kay$ any Hilbert space of dimension at least~$\dim(\aitch)$, and
$\ket{\phi_i}$ any purifications of~$\brho_i$ in $\aitch \tensor
\kay$.
Then, there is a local unitary transformation~$U$ on~$\kay$ that
maps~$\ket{\phi_2}$ to~$\ket{\phi'_2} = I\tensor U\ket{\phi_2}$ such
that
$$\trn{ \ketbra{\phi_1}{\phi_1} - \ketbra{\phi'_2}{\phi'_2}}
\;\;\le\;\; 2 \trn{\brho_1-\brho_2}^{1\over 2}.$$
\end{theorem}
This may be of significance in cryptographic applications as well.

\section{Preliminaries}
\label{sec-pre}

\subsection{The communication complexity model}

In the quantum communication complexity model~\cite{Y93},
Alice and Bob hold qubits. When the game starts Alice holds a
superposition~$\ket{x}$ and Bob holds~$\ket{y}$ (representing the input
to the two players), and so the
initial joint state is simply~$\ket{x} \tensor \ket{y}$. 
The two parties then play in turns. Suppose it is Alice's turn to play.
Alice can do an arbitrary unitary transformation on her qubits
and then send one or more qubits to Bob. 
Sending qubits does not change the overall superposition, but rather
changes the ownership of the qubits, allowing Bob to 
apply his next unitary transformation on the newly received qubits.
At the end of the protocol, one player makes a measurement and declares
that as the result of the protocol.

In general, each player may also (partially) measure her
qubits during her turn. However, we assume
(by invoking the principle of safe
storage~\cite{BernsteinV97}) that all such measurements
are postponed to the end.  We also assume that the two players do not
modify the qubits holding the input superposition during the protocol.
Neither of these affects the aspect of
communication we focus on in this paper.

The complexity of a quantum (or classical) protocol is the number of qubits 
(respectively, bits) exchanged between the two players.
We say a protocol {\em computes\/}
a function~$f : \X \times \Y \mapsto \{0,1\}$
with~$\epsilon \ge 0$ error, if for any input~$x \in \X,y \in \Y$ 
the probability 
that the two players compute~$f(x,y)$ is at least $1-\epsilon$. 
$Q_\epsilon(f)$ denotes
the complexity of the best quantum 
protocol that computes~$f$ with at most~$\epsilon$ error.

For a player~$P \in \set{{\rm Alice},\;{\rm Bob}}$, 
$Q^{c,P}_{\epsilon}(f)$ denotes
the complexity of the best quantum protocol that
computes~$f$ with at most~$\epsilon$ error 
with only~$c$ messages, where the first message is sent by~$P$.
If the name of the player is omitted from the superscript,
either player is allowed to start the protocol.

We say a protocol~$\cp$ {\em computes\/}~$f$ with~$\epsilon$ error 
with respect to a distribution~$\mu$ on~$\X \times \Y$,
if
$$\prob_{(x,y) \in \mu, \cp}( \cp(x,y)=f(x,y)) \;\;\ge\;\; 1-\epsilon.$$ 
$Q^{c,P}_{\mu,\epsilon}(f)$ is the complexity of 
computing~$f$ with at most~$\epsilon$ error with respect to~$\mu$, 
with only~$c$ messages where the first message is sent by player~$P$.
The following is immediate.

\begin{fact}
For any distribution~$\mu$, number of messages~$c$ and
player~$P$,~$Q^{c,P}_{\mu,\epsilon}(f) \le Q^{c,P}_{\epsilon}(f)$.
\end{fact}

\subsection{Classical entropy and mutual information}

The {\em Shannon entropy\/}~$S(X)$ of a classical random variable~$X$
quantifies the amount of randomness in it. If~$X$
takes values~$x$ in some finite set with probability~$p_x$, its Shannon
entropy is defined as~$S(X) \;=\; - \sum_x p_x \log p_x$.
The {\em mutual information\/}~$I(X:Y)$ of a pair of random variables~$X,Y$
is defined by~$I(X:Y)\;=\;S(X)+S(Y)-S(XY)$. It is a measure of how
correlated the two random variables are.

The following are some basic facts about the mutual information function
that we use in the paper.
For any random variables~$X,Y,Z$,
\begin{eqnarray}
\label{eqn:infeq}
I(X:YZ) & = & I(X:Y)+I(XY:Z)-I(Y:Z) \\
\label{eqn:infineq}
I(X:YZ) & \ge  & I(X:Y).
\end{eqnarray}
Fano's inequality states that if~$Y$ can predict another random
variable~$X$ with an advantage, then~$X$ and~$Y$ have large
mutual information. We use it only in the following simple form.
 
\begin{fact}[Fano's inequality]
\label{cl:mut}
Let~$X$ be a uniformly distributed boolean random variable, and let~$Y$ be a
boolean random variable such that~$\prob(X=Y) \ge {1 \over 2}+\delta$,
where~$\delta \ge 0$. Then~$I(X:Y) \ge 1-H({1\over 2} +\delta)$.
\end{fact}

For other equivalent definitions and properties of these concepts,
we refer the reader to a standard text (such as~\cite{CT91}) on
information theory.
Finally, we give a simple bound on the deviation of the binary entropy
function~$H(p)$ from~$1$ as~$p$ deviates from~$1/2$.

\begin{fact}
\label{cl:delta}
For~$\delta \in [-{1 \over 2}, {1 \over 2}]$, we
have~$H({1 \over 2} + \delta) \;\;\le\;\;  1 - \delta^2$.
\end{fact}
\begin{proof}
From the definition of the binary entropy function, we have
\begin{eqnarray*}
H({1 \over 2} + \delta)
    & = & 1 -{1 \over 2}
          [ (1+2 \delta) \log(1+2\delta) + (1-2\delta) \log(1-2\delta) ].
\end{eqnarray*}
Using the expansion~$\ln(1+x) \;\;=\;\; x - \frac{x^2}{2} +
\frac{x^3}{3} - \frac{x^4}{4} + \cdots$ for~$\size{x} < 1$, and
simplifying, we get
\begin{eqnarray*}
H({1 \over 2} + \delta)
  & = &
    1 - (\log {\rm e}) \left[ \left( 2 - \frac{2^2}{2\cdot 2} \right) \delta^2
    + \left( \frac{2^3}{3} - \frac{2^4}{2\cdot 4} \right) \delta^4
    + \left( \frac{2^5}{5} - \frac{2^6}{2\cdot 6} \right) \delta^6
    + \cdots \right] \\
  & \le & 1 - \delta^2,
\end{eqnarray*}
which is the claimed bound.
\end{proof}

\subsection{The density matrix and the trace norm}

The quantum mechanical analogue of a random variable is 
a probability distribution over superpositions, also
called a {\em mixed state}. Consider the mixed state~
$X = \{p_i,\ket{\phi_i}\}$, where the
superposition~$\ket{\phi_i}$ is drawn with probability~$p_i$. 
The {\em density matrix} of the mixed state $X$ is 
~$\brho_X  = \sum_i p_i \ketbra{\phi_i}{\phi_i}$. 
The following properties of density matrices are immediate from the
definition: every density matrix~$\brho$ 
is Hermitian, i.e., $\brho = \brho^\dagger$, has 
unit trace, i.e., $\trace(\brho) = \sum_i \brho(i,i) = 1$, and
is positive semi-definite, i.e., $\bra{\psi}\brho\ket{\psi} \ge 0$
for all~$\ket{\psi}$.
Thus, every density matrix is {\em unitarily diagonalizable\/} and has
non-negative real eigenvalues that sum up to~$1$.

Given a quantum system in a mixed state with density matrix~$\rho$
and a (general) measurement~$\co$ on it,
let~$\rho^{\co}$ denote the classical distribution 
on the possible results that we get by measuring~$\rho$
according to~$\co$. Suppose that it is
some classical distribution~$p_1,\ldots,p_k$
where we get result~$i$ with probability~$p_i$.
Given two different mixed states, we can ask how well one can 
distinguish between the two mixtures, or equivalently, how different the
distributions resulting from a measurement may be.
To quantify this, we consider the~$\ell_1$
metric: if~$p = (p_1,\ldots,p_k)$ and~$q = (q_1,\ldots,q_k)$ are
two probability distributions over~$\{1,\ldots,k\}$,
then the~$\ell_1$ distance between them is~$\norm{p-q}_1
= \sum_i \size{p_i-q_i}$.
A fundamental theorem about distinguishing density matrices
(see~\cite{AKN98}) tells us:

\begin{theorem}
\label{thm:akn}
Let~$\rho_1,\rho_2$ be two density matrices on the same space~$\ch$.
Then for any (general) measurement~$\co$
$$\norm{\rho_1^\co - \rho_2^\co}_1 \;\;\le\;\;
    \trace\,{\sqrt{A^\dagger A}}, $$
where~$A=\rho_1-\rho_2$.
Furthermore, the bound is tight, and the
orthogonal measurement~$\co$ that projects a state on the 
eigenvectors of~$\rho_1-\rho_2$ achieves this bound.
\end{theorem}

Theorem~\ref{thm:akn} shows that the 
density matrix captures all the accessible information that a quantum
state contains. If two different mixtures have the same density matrix (which
is indeed possible) then even though 
they are two distinct distributions,
they are physically, and thus from a computational point of view,
indistinguishable.
As the behavior of a mixed state is completely characterized by its
density matrix we often
identify a mixed state with its density matrix.

The quantity~$\trace\,{\sqrt{A^\dagger A}}$ is of independent interest.
(Note that this is compact notation for the sum of the
(magnitudes of the) singular values of~$A$.)
If we define~$\trn{A} = \trace\,{\sqrt{A^\dagger A}}$
then~$\trn{\cdot}$ defines a norm (the {\em trace norm}),
and has some additional properties
such as~$\trn{A \tensor B} = \trn{A} \cdot \trn{B}$,
$\trn{A}=1$ for any density matrix~$A$ and~$\trn{AB},\trn{BA}
\le \trn{A} \cdot \trn{B}$. (See~\cite{AKN98} for more details.)
We single out the following fact for later use.

\begin{fact}
\label{fac-trn}
If~$\ket{\phi_1},\ket{\phi_2}$ are two pure states,
and~$\rho_i$ is the density matrix of~$\ket{\phi_i}$, then
$$\trn{\rho_1-\rho_2} \;\;=\;\; 2 \sqrt{1-|\la \phi_1 | \phi_2 \ra|^2}.$$
\end{fact}

\subsection{The fidelity measure}

A useful alternative to the trace metric as a measure of closeness of
density matrices is {\em fidelity}, which is defined in terms of the
pure states that can give rise to those density matrices.
A {\em purification\/} of a mixed state~$\brho$ with support in a Hilbert
space~$\aitch$ is any pure state~$\ket{\phi}$ in an extended Hilbert
space~$\aitch \tensor \kay$ such that~$\trace_{\kay}
\ketbra{\phi}{\phi} = \brho$.
Given two density matrices~$\brho_1,\brho_2$ 
on the same Hilbert space~$\aitch$, their {\em fidelity\/} is defined as
$$
F(\brho_1,\brho_2) \;\;=\;\; \sup\, \size{\braket{\phi_1}{\phi_2}}^2,
$$
where the supremum is taken over all purifications~$\ket{\phi_i}$
of~$\brho_i$ in the same Hilbert space~\cite{Jozsa94}.
We state a few properties of this measure:
$0\le F(\brho_1,\brho_2)\le 1$,
$F(\brho_1,\brho_2)=1 \Longleftrightarrow \brho_1=\brho_2$ and
if~$\brho_1 = \density{\phi_1}$, then we have~$F(\brho_1,\brho_2)
= \bra{\phi_1}\brho_2\ket{\phi_1}$.
Jozsa~\cite{Jozsa94} proved that the optimum is always achieved when
finite dimensional density matrices are considered.

\begin{theorem}[Jozsa]
\label{thm-finite-fid}
Let~$\brho_1,\brho_2$ be any two mixed states with support in a finite
dimensional Hilbert
space~$\aitch$,
$\kay$ a Hilbert space of dimension at least~$\dim(\aitch)$,
and~$\ket{\phi_1}$ any purification of~$\brho_1$ in~$\aitch
\tensor \kay$. 
Then there exists a purification~$\ket{\phi_2} \in \aitch
\tensor \kay$ of~$\brho_2$ such that~$\size{\braket{\phi_1}{\phi_2}}^2 =
F(\brho_1,\brho_2)$.
\end{theorem}
Jozsa~\cite{Jozsa94} also gave a simple proof (again for the finite
dimensional case) of the following remarkable 
equivalence first established by Uhlmann~\cite{Uhlmann76}.
$$
F(\brho_1,\brho_2) \;\;=\;\; \left[\trace\left(
                             \sqrt{\brho_1}\,\brho_2\sqrt{\brho_1}
                             \right)^{1\over 2} \right]^2
                   \;\;=\;\; \trn{ \sqrt{\brho_1}\sqrt{\brho_2}}^2.
$$

Using this equivalence, Fuchs and van de Graaf~\cite{FG99}
show that the fidelity and the trace measures of distance between density
matrices are closely related. They prove:

\begin{theorem}[Fuchs, van de Graaf]
\label{thm-fidelity}
For any two mixed states $\brho_1,\brho_2$,
$$
1 - \sqrt{F(\brho_1,\brho_2)} \;\; \le \;\; 
\frac{1}{2} \trn{\brho_1 - \brho_2} \;\;\le\;\;
 \sqrt{1 - F(\brho_1,\brho_2)}.
$$
\end{theorem}

\subsection{Von Neumann entropy and quantum mutual information}

As mentioned earlier, the eigenvalues of a density matrix
are all real, non-negative and sum up to one. Thus, they induce a
probability distribution on the corresponding 
eigenvectors. Since the eigenvectors are all orthogonal, this is
essentially a {\em classical\/} distribution. Every mixed
state with the same density matrix is physically
equivalent to such a canonical classical distribution. It is thus natural
to define the entropy of a mixed state as the {\em Shannon entropy\/}
of this distribution.
Formally, the {\em von Neumann entropy\/}~$S(\brho)$ of a density
matrix~$\brho$ is defined as~$S(\brho) = - \sum_i \lambda_i \log
\lambda_i$,   
where~$\{\lambda_i\}$ is the multi-set of all the eigenvalues
of~$\brho$. More compactly,~$S(\brho) = - \trace\,\brho\log\brho$.

Not all properties of
classical Shannon entropy carry over to the quantum case. For
example it is quite possible that~$S(XY) < S(X)$ as can be 
seen by considering the pure
state~${1 \over \sqrt{2}} (\ket{0}_X\ket{0}_Y+\ket{1}_X\ket{1}_Y)$.
Nonetheless, some of the classical properties do carry over, e.g.,
$S(X) \ge 0$, $S(X)$ is concave and~$S(XY) \le S(X)+S(Y)$.
A property of interest to us is the following, which also
generalizes a classical assertion.

\begin{fact}
\label{cl:cond}
Suppose a quantum system~$A$ is in mixed state~$\set{p_i,\ket{i}}$,
where~$\set{\ket{i}}$ are orthogonal, and~$\bsigma_i$ are
density matrices for another system~$B$, then
$S(\sum_i \, p_i \,\ketbra{i}{i} \tensor \bsigma_i) \;\;
 =\;\; H(A)+\sum_i\, p_i\, S(\bsigma_i)$.
\end{fact}

The density matrix corresponding to a mixed state with superpositions
drawn from a Hilbert space~$\aitch$ is said to have {\em support\/}
in~$\aitch$. 
A density matrix with support in a Hilbert space of
dimension~$d$, 
has~$d$ eigenvalues, hence the entropy
of any such distribution is at most~$\log d$. 
A pure-state has zero entropy. Measuring a pure-state may result in 
a non-trivial mixture and positive entropy. In general,
orthogonal measurements increase the entropy.
For a comprehensive introduction to this concept and
its properties see, for instance,~\cite{P98}.

We define the ``mutual information''~$I(X:Y)$ 
of two disjoint systems $X,Y$ in analogy with classical mutual
information:
$I(X:Y)=S(X)+S(Y)-S(XY)$,
where~$XY$ is density matrix of the system that includes the qubits
of both systems.
Again, not all properties of
classical mutual information carry over to the quantum case. For
example, it is not true in general that~$I(X:Y) \le S(X)$. Nonetheless,
some of the intuition we have about mutual information still applies.
Equation~(\ref{eqn:infeq}) still holds, as follows immediately
from the definition. Equation~(\ref{eqn:infineq})
also continues to be true, but its proof is much more involved. 
It is in fact equivalent to the {\em strong sub-additivity property\/}
of von Neumann entropy.
An important consequence of this property is that local measurements
can only decrease the amount of mutual information. A special case of
this is
the classic Holevo theorem~\cite{H73} from quantum information 
theory, which bounds the amount of information we can extract from a quantum 
encoding of classical bits.

\begin{theorem}[Holevo]
\label{thm-holevo}
Let~$x \mapsto \bsigma_x$ be any quantum encoding of bit strings into
density matrices.
let~$X$ be a random variable with a distribution given by~$\prob(X=x) =
p_x$, let~$Q$ be the quantum encoding of~$X$ according to this map,
and let~$\bsigma = \sum_x p_x \bsigma_x$.
If~$Y$ is any random variable obtained by
performing a measurement on the encoding, then
$$
I(X:Y) \;\; \le \;\; I(X:Q)
\;\;=\;\; S(\bsigma) - \sum_x p_x S(\bsigma_x).
$$
\end{theorem}

In analogy with classical conditional entropy, we define~$S(Y|X) =
\sum_x p_x S(\bsigma_x)$, where~$X$ is a classical random variable
and~$Y$ is a quantum encoding of it given by~$x \mapsto \bsigma_x$. 
We similarly define conditional von Neumann entropy and mutual
information with respect to a classical event. Thus, for
example,~$I(X:Y) = S(Y) - S(Y|X)$.

\section{The average encoding theorem}
\label{sec-tech}

The average encoding theorem asserts that if a quantum encoding has little
correlation with the encoded classical information then the encoded
states are essentially indistinguishable. In particular, they are all
``close'' to the {\em average\/} encoding. This theorem formalizes a
very intuitive idea and might seem to be immediate from Holevo's
theorem. 
However, there is a subtle difference: 
in Holevo's theorem one is interested in a {\em single\/} measurement that
{\em simultaneously} distinguishes all the states,
whereas in our case we are interested in the 
{\em pairwise\/} distinguishability of the encoded states.
We first prove:

\begin{theorem}
\label{thm:avgdist}
Let~$x \mapsto \bsigma_x$ be a quantum encoding
mapping~$m$ bit strings~$x\in \set{0,1}^m$ into mixed
states~$\bsigma_x$.
Let~$X$ be distributed uniformly over~$\set{0,1}^m$
and let~$Q$ be the
encoding of~$X$ according to this map.
Denote 
$\Delta={1 \over 2^{2m}} 
\sum_{x_1,x_2 \in \set{0,1}^m} \trn{\bsigma_{x_1}-\bsigma_{x_2}}$.
Then $I(X:Q) \ge 1-H({1 + \Delta \over 2})$.
\end{theorem}

\begin{proof}
We start with the special case of~$m=1$.
By Theorem~\ref{thm:akn},
there is a measurement~${\cal O}$ on~$Q$ that realizes the trace norm
distance~$t=\trn{\bsigma_0-\bsigma_1}$ between~$\bsigma_0$ and~$\bsigma_1$.
Using Bayes' strategy (see, for example,~\cite{FG99}), the
resulting distributions can be distinguished with probability~${1 \over
2}+{t \over 4}$.
Let~$Y$ denote the classical random variable holding the result
of this entire procedure. 
We have~$\prob(Y=X) = {1 \over 2}+{t \over 4}$.
Thus, by Fano's Inequality,
$$
I(X:Y) \;\; \ge \;\; 1-H({1 \over 2}+{t \over 4})
$$
We complete the proof for~$m = 1$ by noticing that measurements can only
reduce the entropy, hence~$I(X:Q) \ge I(X:Y)$, and
that~$\Delta={t \over 2}$.

To prove the theorem for general~$m$ we reduce it to the~$m=1$ case.
We do this by partitioning the set of strings into pairs with ``easily''
distinguishable encoding.

\begin{lemma}
\label{lem:pairing}
There is a set of~$2^m/2$ disjoint pairs~$(x_{2i-1},x_{2i})$
which together cover~$\set{0,1}^m$ such that
$$
{2 \over 2^m} \sum_i \trn{\bsigma_{x_{2i-1}}-\bsigma_{x_{2i}}}
\;\;\ge\;\; \Delta.
$$
\end{lemma}

\begin{proof}
The expectation of the LHS over a random pairing is~${{2^m}\over {2^m -
1}} \Delta$; hence there is a pairing that achieves this $\Delta$.
\end{proof}

We now fix this pairing. 
Let $Z_i$ denote the set of elements in the~$i$'th pair,
i.e., $Z_i=\set{x_{2i-1},x_{2i}}$
and~$\Delta_i = \trn{\bsigma_{x_{2i-1}}-\bsigma_{x_{2i}}}$.
We know that~${2 \over 2^m} \sum \Delta_i \ge \Delta$. 
Let us also denote $f(\delta)=1-H({1 +\delta \over 2})$.
From the base case $m=1$, we know that for any $i=1,\ldots,2^m/2$,
$I(X:Q\,|\,X\in Z_i) \ge f(\Delta_i)$. 
Thus we get:
$$
S(Q\,|\, X\in Z_i) - {1 \over 2} [S(\bsigma_{x_{2i}}) - S(\bsigma_{x_{2i+1}})]
\;\;\ge\;\; f(\Delta_i).
$$
Averaging all the $2^m/2$ equations yields:
\begin{eqnarray*}
{2 \over 2^m} \sum_i S(Q\,|\,X\in Z_i) - {1 \over 2^m} \sum_x S(\bsigma_x) 
& \ge &
{2 \over 2^m} \sum_i f(\Delta_i)
\end{eqnarray*}
By the concavity of the entropy function,
$S(Q) \ge {2 \over 2^m} \sum_i S(Q\, |\, X\in Z_i)$,
and by definition
${1 \over 2^m} \sum_x S(\bsigma_x) = S(Q|X)$.
Therefore,
$$
I(X:Q) \;\; = \;\; S(Q) - S(Q|X) \;\;\ge\;\; {2 \over 2^m} \sum_i f(\Delta_i).
$$
Since~$f$ is convex,
${2 \over 2^m} \sum_i f(\Delta_i) \ge f({2 \over 2^m} \sum_i \Delta_i)$.
Also,~$f(\delta)$ is monotone increasing for~$0 \le \delta \le {1\over
2}$, so~$f({2 \over 2^m} \sum_i \Delta_i) \ge f(\Delta)$.
Together this yields~$I(X:Q) \ge f(\Delta)$, as required.
\end{proof}

Now, we can easily deduce Theorem~\ref{thm-mi}.

\begin{proofof}{Theorem~\ref{thm-mi}}
Let~$\Delta'={1 \over 2^m} \sum_{x_1} \trn{\bsigma_{x_1} -\bsigma}$.
We have:
\begin{eqnarray*}
\Delta' ~=~
{1 \over 2^m} \sum_{x_1} \trn{\bsigma_{x_1} -\bsigma} 
& = & 
{1 \over 2^m} \sum_{x_1} 
 \trn{ {1 \over 2^m} \sum_{x_2} (\bsigma_{x_1} -\bsigma_{x_2}) } 
 ~ \le ~
{1 \over 2^{2m}} \sum_{x_1,x_2} \trn{\bsigma_{x_1} -\bsigma_{x_2}} 
~\le~ \Delta 
\end{eqnarray*}

By Theorem~\ref{thm:avgdist}, $I(X:Q) \ge 1-H({{1+\Delta}\over 2})$, and
by Fact~\ref{cl:delta} we have~$1-H({1+\Delta \over 2})
\ge 1-(1-({\Delta \over 2})^2)=
{\Delta^2 \over 4}$. Thus,~$\Delta' \le \Delta \le 2\sqrt{I(X:Q)}$.  
\end{proofof}


\section{Local transition between bipartite states}
\label{sec-local}

Lo and Chau~\cite{LC98} and Mayers~\cite{M97} proved:

\begin{theorem}[Lo and Chau; Mayers]
\label{thm:LC}
Suppose~$\ket{\phi_1}$ and~$\ket{\phi_2}$
are two pure states in the Hilbert space $\aitch \tensor \kay$,
such that~$\trace_\kay\ketbra{\phi_2}{\phi_2} =
\trace_\kay\ketbra{\phi_1}{\phi_1}$, i.e.,
the reduced density matrix of~$\ket{\phi_2}$ to~$\aitch$
is the same as the reduced density matrix of~$\ket{\phi_1}$ to~$\aitch$.
Then, there is a local unitary transformation~$U$ on~$\kay$ such 
that~$I\tensor U \ket{\phi_2} = \ket{\phi_1}$.
\end{theorem}
The theorem follows by examining the Schmidt 
decomposition~\cite{P98} of the two states.

A natural generalization of this is to the case where 
the reduced density matrices are close to each other but not quite
the same, which is what appears in Theorem~\ref{thm:close}.
Lo and Chau~\cite{LC98} and Mayers~\cite{M97} considered this case as well.
Theorem~\ref{thm:close} follows from their work by using
the newer results of~\cite{FG99} stated in Theorem~\ref{thm-fidelity}.

\begin{proofof}{Theorem~\ref{thm:close}}
By Theorem~\ref{thm-finite-fid}, there exists a 
purification~$\ket{\phi'_2} \in \aitch \tensor \kay$ of~$\brho_2$ such
that~$\size{\braket{\phi_1}{\phi'_2}}^2 = F(\brho_1,\brho_2)$.
Since~$\ket{\phi_2}$ and~$\ket{\phi'_2}$ have the same reduced density
matrix in~$\aitch$, by Theorem~\ref{thm:LC}, there is a (local) unitary
transformation~$U$ on~$\kay$ such that~$I\tensor U \ket{\phi_2} =
\ket{\phi'_2}$. Moreover, by Fact~\ref{fac-trn} we have
$$
\trn{\density{\phi_1}-\density{\phi'_2}} 
    \;\; = \;\; 2 \sqrt{1-\size{\braket{\phi_1}{\phi'_2}}^2}
    \;\; = \;\; 2 \sqrt{1-F(\brho_1,\brho_2)}.
$$
By Theorem~\ref{thm-fidelity},~
$\sqrt{F(\brho_1,\brho_2)} \;\ge\;
1-\frac{1}{2} \trnorm{\brho_1- \brho_2}$, so
$$
1 - F(\brho_1,\brho_2)
    \;\; \le \;\;
    1 - \left(1 - \frac{1}{2} \trnorm{\brho_1- \brho_2}\right)^2
    \;\; \le \;\; \trnorm{\brho_1-\brho_2}.
$$
This, when combined with the earlier bound on the trace distance
between~$\ket{\phi_1},\ket{\phi'_2}$ gives us the required result.
\end{proofof}


\section{The role of interaction in quantum communication}

In this section, we prove that allowing more interaction between two
players in a quantum communication game can substantially reduce the
amount of communication required. We first define a communication
problem and state
our results formally (giving an overview of the proof), and then give the
details of the proofs.

\subsection{The communication problem and its complexity}

In this section, we give the main components of the proof of
Theorem~\ref{thm-main}.
We define a sequence of problems~$S_0,S_1,\ldots,S_k,\ldots$ by induction.
The problem~$S_1$ is the index function, i.e.,
Alice has a~$n$-bit string~$x \in \X_1 = \set{0,1}^n$,
Bob has an index~$i \in \Y_1 = [n]$ and the desired output is~$S_1(x,i)=x_i$.  
Suppose we have already defined the
function~$S_{k-1}: \X_{k-1} \times  \Y_{k-1} \to \set{0,1}$.
In the problem~$S_{k}$, Alice has as input her part of~$n$
independent instances of~$S_{k-1}$, i.e., $x \in \X_{k-1}^n$,
Bob has his share of~$n$ independent instances of~$S_{k-1}$, i.e.,
$y \in \Y_{k-1}^n$, and
in addition, there is an extra input~$a \in [n]$ 
which is given to Alice if~$k$ is even and to Bob if~$k$ is odd.
The output we seek is the solution to the~$a$th instance of~$S_{k-1}$.
In other words, $S_{k}(x_1,\ldots,x_n,a,y_1,\ldots,y_n)=S_{k-1}(x_a,y_a)$.

Note that the input size to the problem~$S_k$ is~$N = \Theta(n^k)$.
If we allow~$k$ message exchanges for solving the problem, it can be
solved by exchanging~$\Theta(\log N) = \Theta(k\log n)$ bits: for~$k =
1$, Bob sends Alice the index~$i$ and Alice then knows the answer;
for~$k > 1$, the player with the index~$a$ sends it to the other player
and then they recursively solve for~$S_{k-1}(x_a,y_a)$. However, we show
that if we allow one less message, then no quantum protocol can
compute~$S_k$ as efficiently. In fact, no quantum protocol can compute
the function as efficiently even if we require small probability of
error only on average.

\begin{theorem}
\label{thm-lb}
For all constant~$k \ge 1$, $0 \le \epsilon < {1\over 2}$,~~
$
Q^k_{U,\epsilon}(S_{k+1})
    \;\;\ge\;\; \Omega\left( N^{1/(k+1)} \right).
$
\end{theorem}
In fact, we prove a stronger intermediate claim. Let~$P_1$ be Alice, and 
for~$k \ge 2$, let~$P_k$ denote the player
that holds the index~$a$ in an instance of~$S_k$
($a$ indicates which of the~$n$ instances of~$S_{k-1}$ to solve).
Let~$\bar{P}_k$ denote the other player. We refer to~$\bar{P}_k$
as the ``wrong'' player to start a protocol for~$S_k$. The stronger
claim is that any~$k$ message protocol for~$S_k$ 
in which the wrong player starts is exponentially inefficient as
compared to the~$\log N$ protocol described above.

\begin{theorem}
\label{thm-lbi}
For all constant~$k \ge 1$, $0 \le \epsilon < {1\over 2}$,~~
$
Q^{k,\bar{P}_k}_{U,\epsilon}(S_k)
    \;\;\ge\;\; \Omega\left( N^{1/k} \right).
$
\end{theorem}
In fact, there is a classical~$k$-message protocol in which the wrong
player starts with complexity~$O(n)$, so our lower bound is optimal.

Theorem~\ref{thm-lb} now follows directly.

\begin{proofof}{Theorem~\ref{thm-lb}}
It is enough to show the lower bound for the two cases when the protocol
starts either with~$P_{k+1}$ or with the other player.

Let~$P_{k+1}$ be the player to start.
Note that if we set~$a$ to a fixed value, say~$1$, then we get an
instance of~$S_k$. So~$Q^{k,P_{k+1}}_{U,\epsilon}(S_k) \le
Q^{k,P_{k+1}}_{U,\epsilon}(S_{k+1})$. But~$P_{k+1} = \bar{P}_k$, so
the bound of Theorem~\ref{thm-lbi} applies.

Let player~$\bar{P}_{k+1}$ be the one to start. Then, observe that if we
allow one more message (i.e.,~$k+1$ messages in all), the complexity of
the problem only decreases: $Q^{k+1,\bar{P}_{k+1}}_{U,\epsilon}(S_{k+1})
\le Q^{k,\bar{P}_{k+1}}_{U,\epsilon}(S_{k+1})$. So we again get the same
bound from Theorem~\ref{thm-lbi}.
\end{proofof}

We prove Theorem~\ref{thm-lbi} by induction.
First, we show that the index function is hard to 
solve with one message if the wrong player starts. 
This essentially follows from the lower bound for random access codes
in~\cite{Nayak99}.
The only difference is that
we seek a lower bound for a protocol that has low error probability 
{\em on average\/} rather than in the worst case, so we need
a refinement of the original argument. We give this in the next section.

\begin{lemma}
\label{lem:k=1}
For any~$0 \le \epsilon \le 1$,~~$Q^{1,A}_{U,\epsilon}(S_1) \;\;\ge\;\;
(1-H(\epsilon))n$.
\end{lemma}

\remove{
\begin{proof}
Let~$Q$
denote the message sent by Alice. 
On the one hand, $I(Q:X) \le \ell$, the number of
bits in~$Q$. On the other hand,
for~$y \in \set{0,1}^j$, let~$\epsilon_y$ be the error probability
when~$x_l = y_l$, $l \le j$, and the index~$i = j+1$. Note
that~
$\epsilon = {1 \over n}\sum^{n-1}_{j = 0} {1 \over {2^j}} 
\sum_{y\in \set{0,1}^j} \epsilon_y$. We have~$I(Q_y:X_{j+1}) \ge 1 -
H(\epsilon_y)$.
By Lemma~\ref{cl:together},
$$
I(Q:X) \;\;\ge\;\;
\sum^{n-1}_{j = 0} {1 \over {2^j}}
\sum_{y \in \set{0,1}^j} I(Q_y:X_{j+1}) \;\;\ge\;\; 
(1 - H(\epsilon))n,
$$
using the concavity of the entropy function.
\end{proof}
}

Next, we show that if we can solve~$S_k$ with~$k$ messages with the
wrong player starting, then we can also solve~$S_{k-1}$ with 
only~$k-1$ messages of almost the same total length, again with the wrong
player starting, at the cost of a slight increase in the
average probability of error.

\begin{lemma}
\label{lem:reduction}
For all~$k \ge 2$, $0 \le \epsilon < {1\over 2}$,~~
$
Q^{k-1,\bar{P}_{k-1}}_{U,\epsilon'}(S_{k-1})
\;\le\; \ell + \log n,
$
where~$\ell = Q^{k,\bar{P}_k}_{U,\epsilon}(S_{k})$,
and~$\epsilon'=\epsilon+4(\ell/n)^{1/4}$.
\end{lemma}
We defer the proof of this lemma to a later section, but show how it
implies Theorem~\ref{thm-lbi} above.

\begin{proofof}{Theorem~\ref{thm-lbi}}
We prove the theorem by induction on~$k$. 
The case~$k=1$ is handled by Lemma~\ref{lem:k=1}.
Suppose the theorem holds for~$k - 1$. We prove by contradiction that
it holds for~$k$ as well.

If~$Q^{k,\bar{P}_k}_{U,\epsilon}(S_{k}) = o(n)$, then by
Lemma~\ref{lem:reduction} there is a~$k-1$ message 
protocol for~$S_{k-1}$ with the wrong
player starting, with error~$\epsilon' = \epsilon + o(1) < {1\over 2}$,
and with the same communication complexity~$o(n)$. This
contradicts the induction hypothesis.
\end{proofof}

\subsection{Hardness of the index function}

We now prove the average case hardness of the index function. 

\begin{proofof}{Lemma~\ref{lem:k=1}}
Let~$Q$ denote the message sent by Alice. 
For a prefix~$y \in \set{0,1}^i$ of length~$i \ge 0$,
let~$Q_y$ be the encoding which is prepared by first
fixing~$x_1=y_1,\ldots,x_i=y_i$
and then choosing~$x_{i+1},\ldots,x_m$ at random and sending
the state~$\bsigma_x$. Its density matrix is given by
$$
\bsigma_y \;\;=\;\; {1 \over 2^{m-i}} \sum_{z \in \set{0,1}^{m-i}}
                    \bsigma_{yz}.
$$

On the one hand,~$I(Q:X) \le \ell$, the number of
qubits in~$Q$. On the other hand,
for~$y \in \set{0,1}^j$, let~$\epsilon_y$ be the error probability
when~$x_l = y_l$, $l \le j$, and the index~$i = j+1$. Note
that~
$\epsilon = {1 \over n}\sum^{n-1}_{j = 0} {1 \over {2^j}} 
\sum_{y\in \set{0,1}^j} \epsilon_y$. Moreover, 
we have~$I(Q_y:X_{j+1}) \ge 1 - H(\epsilon_y)$, since Bob has a
measurement that predicts~$X_{j+1}$ with probability~$1-\epsilon_y$
given~$Q_y$.
We now claim that
\begin{lemma}
\label{cl:together}
${1 \over 2^m} \sum_x \sum_{i=0}^{m-1} I(Q_{x_1\cdots x_i}:X_{i+1})
 \;\;\le\;\; I(Q:X)$.
\end{lemma}
By this lemma,
$$
I(Q:X) \;\;\ge\;\;
\sum^{n-1}_{j = 0} {1 \over {2^j}}
\sum_{y \in \set{0,1}^j} I(Q_y:X_{j+1}) \;\;\ge\;\; 
(1 - H(\epsilon))n,
$$
using the concavity of the entropy function.
\end{proofof}

\begin{proofof}{Lemma~\ref{cl:together}}
By the definition of mutual information, and using Fact~\ref{cl:cond},
\begin{eqnarray*}
I(Q X_1 \cdots X_i : X_{i+1})
    & = & S(Q X_1 \cdots X_i) + S(X_{i+1}) - S(Q X_1 \cdots X_{i+1}) \\
    & = & 
   [ i + \frac{1}{2^i} \sum_{y \in \set{0,1}^i} S(\bsigma_y) ]
 + [1] 
 - [ (i+1) + \frac{1}{2^{i+1}} \sum_{y \in \set{0,1}^{i+1}} S(\bsigma_y) ] \\
    & = & \frac{1}{2^i} \sum_{y \in \set{0,1}^i} [ S(\bsigma_y) -
          {1\over 2} (S(\bsigma_{y0}) + S(\bsigma_{y1})) ] \\
    & = & \frac{1}{2^i} \sum_{y \in \set{0,1}^i} I(Q_y:X_{i+1}).
\end{eqnarray*}
Moreover, from Properties~(\ref{eqn:infeq}) and~(\ref{eqn:infineq}), 
\begin{eqnarray*}
I(Q:X) & \ge   & \sum_{i=0}^{m-1}  I(Q,X_1,\ldots,X_i:X_{i+1}) \\
       & = & \sum_{i=0}^{m-1}  
               \frac{1}{2^i} \sum_{y \in \set{0,1}^i} I(Q_y:X_{i+1}) \\
       & =   & \sum_{i=0}^{m-1}  
  \frac{1}{2^m} \sum_{y \in \set{0,1}^m} I(Q_{y_1\cdots y_i}:X_{i+1}),
\end{eqnarray*}
which proves the claim.
\end{proofof}

\subsection{The reduction step}

In this section, we show how an efficient protocol for~$S_k$ gives rise
to an efficient protocol for~$S_{k-1}$. The gross 
structure of the argument is the same as in~\cite{MNSW95,KN97}.
However,
we use entirely new techniques from quantum information theory, as
developed in Section~\ref{sec-tech} and~\ref{sec-local} and also get 
better bounds in the process.

\begin{proofof}{Lemma~\ref{lem:reduction}}
For concreteness, we assume that~$k$ is even, so that~$\bar{P}_k$ is
Bob.
Let~$\cp$ be a protocol that solves~$S_{k}$ with respect to~$U$
with~$\ell$ message qubits, error~$\epsilon$,
and~$k$ messages starting with Bob.
We would like to concentrate on inputs where~$a$ is fixed to a
particular value in~$[n]$. This would give rise to an instance
of~$S_{k-1}$ that is also solved by~$\cp$, but with~$k$ messages.
An easy argument shows the first message carries almost no
information about~$y_a$, and we would like to argue that it is not 
relevant for solving~$S_{k-1}$. 
However, the correctness of the protocol relies on
the message,
so we try to reconstruct the message with {\em Alice\/} starting
the protocol instead.
We give the details below.

We first derive a protocol~$\cp'$ which has low error on 
an input for~$S_k$ generated as below
(we call the resulting distribution~$U_{a=j}$):
$x_1,\ldots,x_n$ are chosen uniformly at random from~$\X_{k-1}$,
$a$ is set to~$j$,
$y_j$ is chosen uniformly at random from~$\Y_{k-1}$, and
for all~$i\not=j$, register~$Y_i$ is initialized to the
state~$\sum_{z\in \Y_{k-1}} \ket{z}$ (normalized).

Let~$\epsilon_j$ denote the error of~$\cp$ with respect
to the distribution~$U_{a=j}$. Note that~${1 \over n} \sum_i 
\epsilon_i \le \epsilon$,
since having the~$Y_i$ in a uniform superposition over all possible
inputs has the same effect on the result of the protocol as having it
randomly distributed over the inputs (recall that we require that the input
registers are not changed during a quantum protocol).
Let~$\mu_j$ be the mutual information~$I(M:Y_j)$ in the
protocol~$\cp$ when run on the mixed state~$U_{a=j}$ with~$y_j$ being
chosen randomly. 

\begin{lemma}
\label{lem:zero}
There is a protocol~$\cp'$ which solves~$S_k$ with respect to the
distribution~$U_{a=j}$ with error~$\delta_j = \epsilon_j +
4 \mu_j^{1/4}$ error,
$\ell$ message qubits and~$k$ rounds
starting with Bob, such that~$I(M:Y_{j}) = 0$.
\end{lemma}

The protocol~$\cp'$ is obtained by
slightly modifying the first message 
in protocol~$\cp$ so that it
is {\em completely\/} independent of~$Y_j$. This only affects the
average probability of error. 
Moreover,
in~$\cp'$ the first message does not carry any information about~$y_j$ and
is therefore completely independent of it.
Intuitively this means that Alice does not need to get that
message at all, or equivalently that she can recreate it herself. 
This gives a protocol for solving~$S_{k-1}(x_{j},y_{j})$
with~$k-1$ messages and with Alice starting.

\begin{lemma}
\label{lem:simu}
There is a protocol~$\cp''$ that solves~$S_{k-1}$ with respect to~$U$
with~$\epsilon'$ error, $\ell + \log n$ message qubits and~$k-1$ messages
starting with Alice.
\end{lemma}

Together we get $Q^{k-1,A}_{U,\epsilon'}(S_{k-1}) \le \ell + \log n$
as claimed.
\end{proofof}

\subsection{Proof of Lemmas~\ref{lem:zero} and~\ref{lem:simu}}

\begin{proofof}{Lemma~\ref{lem:zero}}
First consider the case when~$Y_j$ is fixed to some~$z$, but the rest of
the inputs are as in~$U_{a=j}$.
In protocol~$\cp$ Bob applies a unitary transformation~$V$ on
his qubits and
computes~$\ket{\phi(z)} = V \ket{\bar{0},Y_1,\ldots,Y_n}$
in register~$M$ (for the message) and~$B$ (for Bob's ancilla and input).
In~$\cp'$ the message computation is slightly different.
Instead of computing~$\ket{\phi(z)}$,
Bob computes~$\ket{\phi'} = V 
\ket{\bar{0},Y_1,\ldots,Y_{j-1}}\ket{\psi}\ket{Y_{j+1},
\ldots,Y_n}$, where~$\ket{\psi}$ is the uniform superposition
over~$\Y_{k-1}$.
Clearly, in~$\cp'$ the state~$\ket{\phi'}$ and hence the message~$M$ 
does not depend on~$y_j = z$, hence~$I(M:Y_j)=0$ when~$y_j$ is chosen
randomly.

Let us denote by~$\rho_M(z)$
the reduced density matrix of the message
register $M$ in~$\cp$ when the input is drawn according to~$U_{a=j}$
and~$y_j=z$, and let the corresponding density matrix for~$\cp'$
be~$\rho_M$.
Clearly, $\rho_M = {1 \over \size{\Y_{k-1}}} \sum_{z \in \Y_{k-1}} \rho_M(z)$.
Let~$t_z= \trn{\rho_M -\rho_M(z)}$.
By Theorem~\ref{thm-mi} we know that~$E_z t_z \le 2\sqrt{ \mu_j }$.

Protocol~$\cp'$ generates the pure state~$\ket{\phi'}$,
while the desired pure state is~$\ket{\phi(z)}$.
Bob, who knows~$y_j=z$
knows both~$\ket{\phi(z)}$ and~$\ket{\phi'}$. 
By Theorem~\ref{thm:close} there is a local 
unitary transformation~$T_z$ acting on register~$B$ alone, such that
\begin{eqnarray*}
\trn{ \density{T_z\phi'} - \density{\phi(z)} } & \le & 2\sqrt{t_z}.
\end{eqnarray*}
The next step in protocol~$\cp'$ is that Bob applies
 the transformation~$T_z$ to his register~$B$. 
After that, protocol~$\cp'$ proceeds exactly as in~$\cp$.
Therefore, for a given~$z$, the probability that~$\cp$ and~$\cp'$
disagree on the result is at most~$2\sqrt{t_z}$, and the 
error probability of~$\cp'$ on~$U_{a=j}$ is at most
$$
\delta_j \;\; = \;\; \epsilon_j +  2 E_z \sqrt{t_z}
         \;\; \le \;\; \epsilon_j + 2 \sqrt{E_z t_z}
         \;\; \le \;\; \epsilon_j + 4 \mu_j^{1/4},
$$
where the second step follows from Jensen's inequality.
\end{proofof}


\begin{proofof}{Lemma~\ref{lem:simu}}
Protocol~$\cp''$ solves an instance of~$S_{k-1}$.
Alice is given an input~$\hat{x} \in_R \X_{k-1}$
and Bob is given an input~$\hat{y} \in_R \Y_{k-1}$.
The protocol proceeds as follows. Alice and Bob first
reduce the problem to an~$S_k$ instance taken
from the distribution~$U_{a=j}$ for a random~$j$. To do that,
Alice picks~$j \in [n]$ at random, sets~$a = j$ and sends it to
Bob;
Alice sets~$x_j=\hat{x}$ and Bob sets~$y_j=\hat{y}$;
Alice picks~$x_i \in_R \X_{k-1}$ for~$i \neq j$; and
Bob initializes each register~$Y_i$ for~$i \neq j$
with~$\sum_{z \in \Y_{k-1}} \ket{z}$ (normalized).

Notice that if Alice and Bob run the protocol~$\cp'$
over this input, then they get the answer~$S_{k-1}(x,y)$
with probability of error at most
$$
\epsilon' \;\; = \;\;  {1\over n} \sum_{i=1}^n \delta_i 
          \;\; \le \;\;  {1 \over n}\sum_{i=1}^n \epsilon_i + 
                        4 {1\over n} \sum_{i=1}^n \mu_i^{1/4}
          \;\; \le \;\;  \epsilon + 4 \left[ {1\over n}
                         \sum_{i=1}^n \mu_i \right]^{1\over 4}.
$$

We claim that
\begin{claim}
\label{cl:j}
$\sum_i \mu_i \le \ell_1$, where~$\ell_1$ is the length of the
message~$M$.
\end{claim}
Hence~$\epsilon' \;\le\; \epsilon + 4 (\ell/n)^{1/4}$.

Alice and Bob do not run the protocol~$\cp'$ itself, but
a modification of it in which Alice sends the first message instead
of Bob, thus reducing the number of rounds to~$k-1$.

Let~$\rho_M$ be the reduced density matrix of register~$M$ holding the
first message that Bob sends to Alice in~$\cp'$, for the input given
above.
By Lemma~\ref{lem:zero}, we know that~$\rho_M$ does not depend
on~$y_j=\hat{y}$.
So~$\rho_M$ is known {\em in advance\/} to Alice.  
Alice starts the protocol~$P''$ by purifying~$\rho_M$. 
More specifically, let~$\set{\ket{e_i}}$ be an eigenvector basis
for~$\rho_M$ with real and positive eigenvalues~$\lambda_i$.
Alice  constructs the superposition~$\sum_i \sqrt{\lambda_i}
\ket{e_i,i}_{MB}$ over two registers~$M$ (containing the eigenvectors)
and~$B$ (containing the index~$i$), and sends register~$B$ to Bob.
The state of the system after this message in~$\cp''$ is
\begin{eqnarray*} 
\ket{\xi} &=& \ket{x_1,\ldots,x_n}_A 
    \tensor \sum_i \sqrt{\lambda_i} \ket{e_i}_M \ket{i}_B 
\end{eqnarray*}
whereas in~$\cp'$ it is
\begin{eqnarray*} 
\ket{\chi(y)} &=& 
    \ket{x_1,\ldots,x_n}_A  \tensor \ket{T_y\phi'}_{MB}.
\end{eqnarray*}
The reduced density matrix of~$\ket{\xi}$ to registers~$AM$
is the same as the reduced density matrix of~$\ket{\chi(y)}$ 
to registers~$AM$.
By Theorem~\ref{thm:LC}, Bob has a {\em local\/} 
unitary transformation~$V_y$ (operating on his register~$B$)
that transforms~$\ket{\xi}$ to $\ket{\chi(y)}$. Bob applies~$V_y$, and 
Alice and Bob then simulate the rest of the protocol~$\cp'$.
From this stage on, the runs of the protocols~$\cp'$ and~$\cp''$ are
identical 
have the same communication complexity and success probability.
\end{proofof}

\begin{proofof}{Claim~\ref{cl:j}}
Note that~$\mu_j$ is
the same as the mutual information~$I(M:Y_j)$ when~$\cp$ is run on the
uniform distribution on~$\X^n_{k-1} \times \Y^n_{k-1}$. So we prove the
claim for the latter.

For any $i$,
$I(Y_i:Y_1 \cdots Y_{i-1} Y_{i+1} \cdots Y_n)=0$.
Therefore
by Properties~(\ref{eqn:infeq}) and~(\ref{eqn:infineq}) (cf.\
Section~\ref{sec-pre})
we have
$$
I(M:Y_1\cdots Y_n) ~\ge~
\sum_{i=1}^n I(M Y_1\cdots Y_{i-1}:Y_i) ~\ge~ \sum_{i=1}^n I(M:Y_i) 
~=~ \sum_i \mu_i
$$ 
As the first message~$M$ contains only~$\ell_1$ qubits,
we have~$\sum_i \mu_i \le I(M:Y_1 \cdots Y_n) \le \ell_1$.
\end{proofof}

\subsection*{Acknowledgements}

We thank Jaikumar Radhakrishnan and Venkatesh Srinivasan for their input
on the classical communication complexity of the pointer jumping and the
subproblem we study in this paper, and Dorit Aharonov for helpful comments.

\bibliographystyle{plain}
\bibliography{bibl}  

\newpage

\end{document}